\documentclass[twocolumn,aps,prb,epsf,showpacs]{revtex4}
\usepackage{amssymb}

\newcommand{\be}{\begin{equation}}
\newcommand{\ee}{\end{equation}}
\newcommand{\bea}{\begin{eqnarray}}
\newcommand{\eea}{\end{eqnarray}}

\begin{document}

\title{Gaussian fluctuation corrections to the BCS mean field gap amplitude
at zero temperature }
\author{\v S. Kos$^{(1)}$, A. J. Millis$^{(2)}$ and A. I. Larkin$^{3}$}
\affiliation{$^{(1)}$Center for Nonlinear Studies, Los Alamos National Laboratory, Los
Alamos, NM 87545\\
$^{(2)}$Department of Physics, Columbia University, 538 W 120th Street, NY,
NY 10027\\
$^{3}$ William I. Fine Theoretical Physics Institute, School of Physics and
Astronomy\\
University of Minnesota, 116 Church Street SE Minneapolis, MN 55455}
\date{\today }

\begin{abstract}
{The leading (Gaussian) fluctuation correction to the weak coupling zero
temperature BCS superconducting gap equation \ is computed. We find that the
dominant contribution comes from the high energies and momenta (compared to
the gap) and gives a correction smaller by the weak-coupling factor $gN_0$
than the mean-field terms. This correction is small due to cancellation of
singular contributions from the amplitude and phase mode at high energies
and momenta. }
\end{abstract}

\maketitle

\section{Introduction}

The Bardeen-Cooper-Schreiffer (BCS) theory of superconductivity \cite{BCS}
is important both as an empirically highly accurate and successful model of
a nontrivial physical phenomenon and as a paradigm for theoretical study of
a wide class of models involving a logarithmically divergent susceptibility.
This paper focusses on the latter aspect. From this point of view the
essence of BCS theory is the observation that in a generic, weakly coupled
fermion system in $d\geq 2$ dimensions, all susceptibilities are
non-negative and remain finite as temperature $T\rightarrow 0$ except for
particle-particle susceptibilities such as 
\begin{eqnarray}
\chi (i\nu ,q) &=&\int d^{d}r\int_{0}^{\beta }d\tau e^{i\nu \tau -i%
\overrightarrow{q}\cdot \overrightarrow{r}}  \label{chi0} \\
&&\left\langle T_{\tau }\left[ \psi _{\downarrow }(\tau ,r)\psi _{\uparrow
}(\tau ,r),\psi _{\uparrow }^{+}(0,0)\psi _{\downarrow }^{+}(0,0)\right]
\right\rangle ,  \nonumber
\end{eqnarray}%
which diverge logarithmically as $q,\nu $ and temperature $T$ tend to $0$.
(By the phrase 'generic' we mean to rule out for example the nesting and van
Hove instabilities occurring in particular models at particular band
fillings).

BCS showed that the logarithmic divergence of $\chi $ signalled the
appearance, at a transition temperature $T_{c},$ of a new order parameter $%
\Delta \sim <\psi _{\downarrow }\psi _{\uparrow }>$. They further argued
that $T_{c}$, the magnitude of $\Delta $ and many other physical
consequences of the ordering could be accurately determined by mean field
theory. The stunning agreement between predictions of the BCS theory and
data on conventional superconductors lends very strong support to this view.

The approach pioneered by BCS has been applied by many workers in many
contexts involving logarithmically or more strongly diverging
susceptibilities; for example to spin and charge density wave instabilities
driven by nesting effects in half filled \ particle-hole symmetric bands 
\cite{Overhauser}, or, in the two dimensional case, at fillings that give
rise to a van-Hove singularity in the density of states \cite{Rice_Scott}.

Some aspects of fluctuations have been understood in detail. Around the time
BCS theory was developed it was understood that in $d<4$ dimensions and for
temperatures sufficiently near to $T_{c}$, nonlinear interactions among very
long wavelength order parameter fluctuations would invalidate a mean field
treatment of the thermally driven transition, and the (rather modest)
temperature window within which conventional superconductors would exhibit
non-mean-field behavior was estimated \cite{Ginsburgcrit}. The effects of
order parameter fluctuations with momenta and energies smaller than the
inverse correlation length and $\Delta $, respectively, on the near-$T_{c}$
properties of conventional superconductors in different dimensions has been
examined in detail \cite{Maki_Thompson} and found, for example, to give a
small negative correction to the mean-field value of the order parameter $%
\Delta $ and in $d=2$ to have a larger effect on $T_{c}$.  Engelbrecht and
co-workers used a functional integral method to investigate the effect of long wavelength
fluctuations on the BCS-Bose-Einstein crossover \cite{Engelbrecht}. Varlamov et al. 
\cite{Varlamov} studied the effect of these long wavelength fluctuations on the
normal state of layered superconductors in the context of high-temperature
superconductivity.

In this paper we investigate corrections to the BCS mean field approximation
due to Gaussian fluctuations. We use a standard functional integral formalism
to study 
the fluctuations
over a wide range of momenta and energies at $T=0$,
where the order parameter is believed to be well developed, and thus the
fluctuations corrections are supposed to be small. Suprisingly, this basic
question seems not to have been addressed in the literature. We calculate
the corrections and find that they are, indeed, small compared to the
mean-field terms. There are three important features: 1. The dominant
correction comes from energies and momenta high compared to the mean-field
gap. 2. The dominant contribution to the correction comes from processes in
which the electrons are scattered nearly parallel to the Fermi surface. 3.
The fluctuation corrections from each of the amplitude and the phase mode
diverge logarithmically at high energies, but with opposite sign so these
divergences cancel leaving a small overall correction.

The rest of this paper is organized as follows: in Section II we present the
formalism to be used. In Section III, we evaluate the polarization kernel
and its derivatives, and use it in Section IV to calculate the fluctuation
correction. Section V is a discussion and conclusion.

\section{General formulae}

We consider for definiteness a model of fermions in $d\geq 2$ spatial
dimensions with energy dispersion $\varepsilon _{p}=p^{2}/2m-\mu $ (as will
become evident below, our results may be trivially generalized to more
realistic dispersions provided nesting and van Hove singularities are
absent). We take the fermions to interact via a short ranged instantaneous
attractive interaction parametrized by a coefficient $g>0$. We follow
Shankar \cite{Shankar} and consider only states within a cutoff $\Lambda
<<p_{F}$ of the fermi surface, expand the density of states $N(\varepsilon
)=\int \frac{d^{d}p}{(2\pi )^{d}}\delta (\varepsilon -\varepsilon _{p})$
around the Fermi energy as 
\begin{equation}
N(\epsilon )=N_{0}+N_{1}\epsilon /v_{F}\Lambda +...,  \label{dosexp}
\end{equation}%
where $N_{0}$ and $N_{1}$ are of the same order of magnitude. The
Hamiltonian thus becomes 
\begin{eqnarray}
H &=&N_{0}\sum_{\sigma }\int d\varepsilon _{p}\varepsilon _{p}c_{p\sigma
}^{+}c_{p\sigma }  \label{H} \\
&&-g\int^{^{\prime }}(dp_{1}...dp_{4})c_{p_{1}\uparrow
}^{+}c_{p_{2}\downarrow }^{+}c_{p_{3}\downarrow }c_{p_{4}\uparrow } 
\nonumber
\end{eqnarray}%
where $(dp)=\frac{d^{d}p}{(2\pi )^{d}}$ and the prime on the integral
indicates that all momenta are restricted to the shell $-\Lambda <\left\vert
p\right\vert -p_{F}<\Lambda $ and that there is a momentum conserving delta
function.

To analyse the model we write it as a functional integral \cite{Eilenberger}%
, decouple the interaction via a Hubbard-Stratonovich transformation and
perform the integral over the fermionic fields obtaining for the partition
function 
\begin{equation}
Z=\int \mathcal{D}\Delta ^{\ast }(\tau ,r)\mathcal{D}\Delta (\tau ,r)e^{S}
\label{Z}
\end{equation}%
with action $S$ given by 
\begin{equation}
S={\mbox{Tr}\ln (-\partial _{\tau }-H(\Delta (\tau ,r)))-\int\limits_{\beta
V}d\tau dr{\frac{|\Delta (\tau ,r)|^{2}}{g}}}.  \label{S}
\end{equation}%
and 
\begin{eqnarray}
{H(\Delta (\tau ,r))} &{=}&N_{0}\sum_{\sigma }\int d\varepsilon
_{p}\varepsilon _{p}c_{p\sigma }^{+}c_{p\sigma }  \label{Hofdel} \\
&&+{\int\limits_{\beta V}d\tau dr}\left( {\Delta (\tau ,r)}\psi _{\uparrow
}^{+}(r,\tau )\psi _{\downarrow }^{+}(r,\tau )+H.c\right)  \nonumber
\end{eqnarray}%
The BCS mean field theory corresponds to a saddle-point approximation to Eq %
\ref{Z}; at the saddle point we have 
\begin{equation}
\Delta (\tau ,r)=\Delta ^{\ast }(\tau ,r)=\Delta _{0},  \label{saddle}
\end{equation}%
and the saddle-point approximation to the action is 
\begin{equation}
S_{0}(\Delta _{0})-S_{0}(\Delta _{0}=0)=\beta V\left( N_{0}\Delta
_{0}^{2}\ln {\frac{2v_{F}\Lambda }{\Delta _{0}}}-{\frac{\Delta _{0}^{2}}{g}%
+...}\right) .  \label{S0}
\end{equation}%
where the ellipsis denotes terms of the order of $\Delta _{0}^{2}$ with
coefficient of order unity, and $v_{F}\equiv d\varepsilon _{p}/dp$ at $p_{F}$
is the Fermi velocity. 
Our calculations are valid in the weak-coupling regime $gN_0<<1$, and these omitted terms are small by at least one power
of $gN_{0}$ relative to terms which we retain. Finally, the saddle point
value of $\Delta _{0}$ is fixed by extremizing $S$ with respect to $\Delta
_{0}$ yielding the familiar BCS gap equation%
\begin{equation}
\frac{1}{g}=N_{0}\left( \ln \frac{v_{F}\Lambda }{\Delta _{0}}+...\right)
\label{BCSgap}
\end{equation}%
where again the ellipsis indicates terms of order unity. \ Thus, as is well
known, within the BCS approximation the $T=0$ gap value $\Delta _{0}$ is
determined only to logarithmic accuracy, i.e. $\ln (\Lambda /\Delta _{0})$
is known up to terms of relative order unity.

To study fluctuation corrections we write: 
\begin{equation}
\Delta (\tau ,r)=\Delta _{0}+\eta (\tau ,r).  \label{Hfluct}
\end{equation}%
We write the fluctuation $\eta $ in terms of its real and imaginary parts 
\begin{equation}
\eta (\tau ,r)=\eta _{1}(\tau ,r)+i\eta _{2}(\tau ,r),
\end{equation}%
which, for the gauge in which $\Delta _{0}$ is real, correspond to the
amplitude and phase fluctuation respectively.

Substitution of Eq \ref{Hfluct} into Eq \ref{Z} and expansion in powers of $%
\delta $ yields $Z=\int D\eta ^{\ast }(\tau ,r)D\eta (\tau
,r)e^{S_{0}+\delta S}$ with 
\begin{eqnarray}
\delta S &=&-\int {\frac{d\nu }{2\pi }}\int\limits^{\Lambda }\left( {\frac{dq%
}{2\pi }}\right) ^{d}\left( {\frac{\delta _{ab}}{g}}+\Pi _{ab}(i\nu
,q)\right)  \label{delS} \\
&&\eta _{a}^{\ast }(i\nu ,q)\eta _{b}(i\nu ,q)+...,  \nonumber
\end{eqnarray}%
where the ellipsis denotes higher powers of $\eta $, $a,b=1.2$ and 
\begin{eqnarray}
\Pi _{ab}(i\nu ,q) &=&\frac{(-1)^{a+b}}{2}\int {\frac{d\omega }{2\pi }}%
\int\limits^{\Lambda }\left( {\frac{dk}{2\pi }}\right) ^{d}  \label{piab} \\
&&Tr[G(i\omega _{+},k_{+})\tau _{a}G(i\omega _{-},k_{-})\tau _{b}]  \nonumber
\end{eqnarray}%
Here $\omega _{\pm }=\omega \pm \frac{\nu }{2}$ and $k_{\pm }=k \pm \frac{q%
}{2}$ while $G$ is the superconducting Green function in Nambu matrix
notation and $\tau _{a},a=1,2$ are the Pauli matrices in particle-hole
space. \ Note that in the instantaneous interaction model it suffices to
place a cutoff on the momentum integral; no frequency{\large \ }cutoff is
required.

We see from Eq \ref{delS} that the propagator corresponding to $\eta $ is of
order $g$ (except at long wavelengths and low energy) so an evaluation of $Z$
via an expansion in powers of $\eta $ yields a series expansion in powers of 
$gN_{0}$ for the free energy and other physical quantities. For example,
restricting the expansion to Gaussian order leads to a correction to the
free energy given by 
\begin{eqnarray}
F_{1} &=&-TS_{1}  \label{fluctuation free energy} \\
&=&VT\sum_{\nu }\int\limits^{\Lambda }\left( {\frac{dq}{2\pi }}\right) ^{d}%
\mbox{Tr}\ln \left( \mathbf{1}+g\mathbf{\Pi }(i\nu ,q)\right)  \nonumber \\
&=&VT\sum_{\nu }\int\limits^{\Lambda }\left( {\frac{dq}{2\pi }}\right) ^{d}{%
\frac{1}{2}}\ln ((1+g\Pi _{0})^{2} \\
&&-g^{2}(\Pi _{1}^{2}+\Pi _{2}^{2}+\Pi _{3}^{2})).  \nonumber
\end{eqnarray}%
Here 
\begin{equation}
\Pi \equiv \Pi _{0}\mathbf{1}+\Pi _{1}\tau _{1}+\Pi _{2}\tau _{2}+\Pi
_{3}\tau _{3}
\end{equation}%
is the polarization matrix written in terms of the Pauli matrices. The
correction to the saddle point equation arising from the Gaussian
fluctuation term is 
\begin{eqnarray}
{\frac{dS_{1}}{d\Delta _{0}^{2}}} &=&-\beta V\int {\frac{d\nu }{2\pi }}%
\int\limits^{\Lambda }\left( {\frac{dq}{2\pi }}\right) ^{d}
\label{fluctgapeq} \\
&&{\frac{{(1+g\Pi _{0})g{\frac{d\Pi _{0}}{d\Delta ^{2}}}-g^{2}(\Pi _{1}{%
\frac{d\Pi _{1}}{d\Delta ^{2}}}+\Pi _{2}{\frac{d\Pi _{2}}{d\Delta ^{2}}}+\Pi
_{3}{\frac{d\Pi _{3}}{d\Delta ^{2}}})}}{(1+g\Pi _{0})^{2}-g^{2}(\Pi
_{1}^{2}+\Pi _{2}^{2}+\Pi _{3}^{2})}}.  \nonumber
\end{eqnarray}%
In the next section, we shall estimate the relative contributions of Eq \ref%
{fluctgapeq} and Eq \ref{BCSgap} to the gap equation.

\section{Calculation of $\Pi $}

This section computes the polarizibilities appearing in Eq \ref{fluctgapeq}.
To simplify notation we work with the action density (i.e. divide by $\beta
V $) and omit the subscript $0$ of $\Delta _{0}$.

Using the explicit mean-field form for $G$, namely 
\begin{equation}
G(i\omega ,k)=-{\frac{i\omega \mathbf{+}\epsilon \tau _{3}+\Delta \tau _{1}}{%
\omega ^{2}+\epsilon ^{2}(k)+\Delta ^{2}}}
\end{equation}%
we find (here $\epsilon _{\pm }=\epsilon _{k\pm q/2}$)

\begin{eqnarray}
\Pi _{0} &=&-\int {\frac{d\omega }{2\pi }}\int\limits^{\Lambda }\left( {%
\frac{dk}{2\pi }}\right) ^{d} \\
&&{\frac{\omega _{+}\omega _{-}+\epsilon _{+}\epsilon _{-}}{\left[ \omega
_{+}^{2}+\epsilon _{+}^{2}+\Delta ^{2}\right] \left[ \omega
_{-}{}^{2}+\epsilon _{-}^{2}+\Delta ^{2}\right] }}  \nonumber \\
\Pi _{2} &=&-\int {\frac{d\omega }{2\pi }}\int\limits^{\Lambda }\left( {%
\frac{dk}{2\pi }}\right) ^{d} \\
&&{\frac{i\omega _{+}\epsilon _{-}-i\omega _{-}\epsilon _{+}}{\left[ \omega
_{+}^{2}+\epsilon _{+}^{2}+\Delta ^{2}\right] \left[ \omega
_{-}{}^{2}+\epsilon _{-}^{2}+\Delta ^{2}\right] }}  \nonumber \\
\Pi _{3} &=&\int {\frac{d\omega }{2\pi }}\int\limits^{\Lambda }\left( {%
\frac{dk}{2\pi }}\right) ^{d} \\
&&{\frac{\Delta ^{2}}{\left[ \omega _{+}^{2}+\epsilon _{+}^{2}+\Delta ^{2}%
\right] \left[ \omega _{-}{}^{2}+\epsilon _{-}^{2}+\Delta ^{2}\right] }}. 
\nonumber
\end{eqnarray}%
and $\Pi _{1}=0$.

First, we analyze $\Pi _{3}$. We linearize the fermion dispersion around the
Fermi surface 
\begin{equation}
\epsilon _{+}=\epsilon +v_{F}{\frac{q}{2}}\mu
\end{equation}%
where $v_{F}$ is the Fermi velocity, and $\mu $ is the cosine of the angle
between $k$ and $q$. For a spherical Fermi surface in $d$ dimensions, $v_{F}$
is a constant, and 
\begin{equation}
\left( {\frac{dk}{2\pi }}\right) ^{d}=\int\limits_{-v_{F}\Lambda
}^{v_{F}\Lambda }N(\epsilon )d\epsilon \int\limits_{0}^{1}{\frac{d\mu (1-\mu
^{2})^{\frac{d-3}{2}}}{\kappa _{d}}};
\end{equation}%
with 
\begin{equation}
\kappa _{d}={\frac{\sqrt{\pi }\Gamma \left( {\frac{d-1}{2}}\right) }{2\Gamma
\left( {\frac{d}{2}}\right) }}.
\end{equation}%
Then 
\begin{equation}
\Pi _{3}(i\nu ,q)=N_{0}\int\limits_{0}^{1}{\frac{d\mu (1-\mu ^{2})^{\frac{d-3%
}{2}}}{\kappa _{d}}}I(\nu ,v_{F}q\mu ,\Delta ;\Lambda )
\end{equation}%
with%
\begin{eqnarray}
I(\nu ,v_{F}q\mu ,\Delta ;\Lambda ) &=&\int \frac{d\omega }{2\pi }%
\int\limits_{-v_{F}\Lambda }^{v_{F}\Lambda }d\varepsilon \\
&&{\frac{\Delta ^{2}}{\left[ \omega _{+}{}^{2}+\epsilon _{+}{}^{2}+\Delta
^{2}\right] \left[ \omega _{-}{}^{2}+\epsilon _{-}{}^{2}+\Delta ^{2}\right] }%
}  \nonumber
\end{eqnarray}%
If $\nu <<v_{F}\Lambda $ and $q<<\Lambda $ then I depends on $\nu ,q$ only
via the combination 
\begin{equation}
r\equiv {\frac{\sqrt{\nu ^{2}+(v_{F}q\mu )^{2}}}{2\Delta }}  \label{rdef}
\end{equation}%
so we may evaluate the $\epsilon $ and $\omega $ integrals at $q=0$ and $\nu
=2\Delta r$ obtaining 
\begin{eqnarray}
I(r;\Lambda ) &\equiv &\int\limits_{-v_{F}\Lambda }^{v_{F}\Lambda }d\epsilon
\int\limits_{-\infty }^{\infty }{\frac{d\omega }{2\pi }} \\
&&{\frac{\Delta ^{2}}{\left[ \left( \omega +\Delta r\right) ^{2}+\epsilon
^{2}+\Delta ^{2}\right] \left[ \left( \omega -\Delta r\right) ^{2}+\epsilon
^{2}+\Delta ^{2}\right] }}  \nonumber \\
&=&{\frac{{1}}{2r\sqrt{1+r^{2}}}}\ln \left( {r}+\sqrt{1+r^{2}}\right) +...
\label{I(r;Lambda)}
\end{eqnarray}%
where the last approximation applies for $r\Delta <<v_{F}\Lambda $ .

Next, to analyze $\Pi _{0}$, we separate $\Pi _{0}(0,0)$ from the formula by
adding and subtracting ${\frac{1}{2}}\left[ \omega _{+}{}^{2}+\epsilon
_{+}^{2}+\Delta ^{2}\right] +{\frac{1}{2}}\left[ \omega _{-}^{2}+\epsilon
_{-}^{2}+\Delta ^{2}\right] $ to get 
\begin{eqnarray}
\Pi _{0}(i\nu ,q) &=&\Pi _{0}(0,0) \\
&&+N_{0}\int\limits_{0}^{1}{\frac{d\mu (1-\mu ^{2})^{\frac{d-3}{2}}}{\kappa
_{d}}}(2r^{2}+1)I(r,\Lambda )  \nonumber
\end{eqnarray}%
For the static uniform polarization, we obtain for $\Delta <<v_{F}\Lambda $ 
\begin{eqnarray}
\Pi _{0}(0,0) &=&-N_{0}\int\limits_{0}^{1}{\frac{d\mu (1-\mu ^{2})^{\frac{d-3%
}{2}}}{\kappa _{d}}} \\
&&\int\limits_{{}}^{{}}{\frac{d\omega }{2\pi }}\int\limits_{-\infty
}^{\infty }d\epsilon {\frac{1}{\omega ^{2}+\epsilon ^{2}+\Delta ^{2}}} 
\nonumber \\
&=&-N_{0}\left( \ln \left( {\frac{v_{F}\Lambda }{\Delta }}\right) +...\right)
\end{eqnarray}%
where again the ellipsis denotes terms of order unity.

Finally, we analyze $\Pi _{2}$. We see that $\Pi _{2}(i\nu ,q)\rightarrow 0\ 
$as$\ \nu \rightarrow 0,$ so 
\begin{equation}
\Pi _{2}=-i\nu J(i\nu ,q).
\end{equation}%
To the accuracy with which we discussed $\Pi _{3}$ and $\Pi _{0}$, we may
say that J is a function of $r$, and we evaluate it at $q=0$, $\nu =2\Delta
r $. Thus, 
\begin{eqnarray}
J &=&\int\limits_{0}^{1}{\frac{d\mu (1-\mu ^{2})^{\frac{d-3}{2}}}{\kappa _{d}%
}}\int\limits_{-v_{F}\Lambda }^{v_{F}\Lambda }d\epsilon \int {\frac{d\omega 
}{2\pi }} \\
&&{\frac{N(\epsilon )\epsilon }{[(\omega +\Delta r)^{2}+\epsilon ^{2}+\Delta
^{2}][(\omega -\Delta r)^{2}+\epsilon ^{2}+\Delta ^{2}]}}.  \nonumber
\end{eqnarray}%
$J$ vanishes in case of particle-hole symmetry, so the constant term in the
density of states (Eq \ref{dosexp}) does not contribute. The $\omega $
integral can be done analytically and we find 
\begin{eqnarray}
J &=&{\frac{N_{1}}{4v_{F}\Lambda }}\int\limits_{0}^{1}{\frac{d\mu (1-\mu
^{2})^{\frac{d-3}{2}}}{\kappa _{d}}} \\
&&\int\limits_{-v_{F}\Lambda }^{v_{F}\Lambda }d\epsilon {\frac{\epsilon ^{2}%
}{\sqrt{\epsilon ^{2}+\Delta ^{2}}(\epsilon ^{2}+\Delta ^{2}+\Delta
^{2}r^{2})}}  \nonumber
\end{eqnarray}

We separate out the logarithmically divergent term and perform the integral
over $\epsilon $ obtaining%
\begin{eqnarray}
J &=&{\frac{N_{1}}{2v_{F}\Lambda }}\int\limits_{0}^{1}{\frac{d\mu (1-\mu
^{2})^{\frac{d-3}{2}}}{\kappa _{d}}} \\
&&\left[ \ln {\frac{v_{F}\Lambda }{\Delta }}-{\frac{\sqrt{r^{2}+1}}{r}}\ln
(r+\sqrt{r^{2}+1})\right]  \nonumber
\end{eqnarray}

The derivatives $d\Pi /d\Delta ^{2}$ needed for Eq \ref{fluctgapeq} may be
computed by straightforward differentiation (recall that $r=\sqrt{\nu
^{2}+(v_{F}q\mu )^{2}}/2\Delta $). We summarize the results: 
\begin{eqnarray}
\Pi _{0} &=&-N_{0}\ln \left( {\frac{v_{F}\Lambda }{\Delta }}\right) + \\
&&N_{0}\int\limits_{0}^{1}{\frac{d\mu (1-\mu ^{2})^{\frac{d-3}{2}}}{\kappa
_{d}}}{\frac{\left( 2r^{2}+1\right) \ln (r+\sqrt{r^{2}+1})}{2r\sqrt{r^{2}+1}}%
}  \nonumber \\
\Pi _{2} &=&N_{1}{\frac{i\nu }{2v_{F}\Lambda }[}-\ln \left( {\frac{%
v_{F}\Lambda }{\Delta }}\right) \\
&&+\int\limits_{0}^{1}{\frac{d\mu (1-\mu ^{2})^{\frac{d-3}{2}}}{\kappa _{d}}}%
{\frac{\sqrt{r^{2}+1}\ln (r+\sqrt{r^{2}+1})}{r}}  \nonumber \\
\Pi _{3} &=&N_{0}\int\limits_{0}^{1}{\frac{d\mu (1-\mu ^{2})^{\frac{d-3}{2}}%
}{\kappa _{d}}}{\frac{\ln (r+\sqrt{r^{2}+1})}{2r\sqrt{r^{2}+1}}} \\
{\frac{d\Pi _{0}}{d\Delta ^{2}}} &=&{\frac{N_{0}}{4\Delta ^{2}}}%
\int\limits_{0}^{1}{\frac{d\mu (1-\mu ^{2})^{\frac{d-3}{2}}}{\kappa _{d}}} \\
&&\left[ {\frac{\ln (\sqrt{r^{2}+1}+r)}{r(r^{2}+1)^{3/2}}}+{\frac{1}{r^{2}+1}%
}\right]  \nonumber \\
{\frac{d\Pi _{2}}{d\Delta ^{2}}} &=&{\frac{N_{1}}{4\Delta ^{2}}}%
\int\limits_{0}^{1}{\frac{d\mu (1-\mu ^{2})^{\frac{d-3}{2}}}{\kappa _{d}}%
\frac{\frac{i\nu }{v_{F}\Lambda }\ln (r+\sqrt{r^{2}+1})}{r\sqrt{r^{2}+1}}} \\
{\frac{d\Pi _{3}}{d\Delta ^{2}}} &=&{\frac{N_{0}}{4\Delta ^{2}}}%
\int\limits_{0}^{1}{\frac{d\mu (1-\mu ^{2})^{\frac{d-3}{2}}}{\kappa _{d}}} \\
&&\left[ {\frac{\left( 2r^{2}+1\right) \ln (\sqrt{r^{2}+1}+r)}{%
r(r^{2}+1)^{3/2}}}-{\frac{1}{r^{2}+1}}\right]  \nonumber
\end{eqnarray}

\section{Evaluation of fluctuation correction.}

This section uses the results of the previous section to evaluate Eq \ref%
{fluctgapeq}. There are two regimes in Eq (\ref{fluctgapeq}): small $r$ ($%
\nu ^{2},(v_{F}q)^{2}\lesssim \Delta ^{2}$) and large $r$. We consider them
in turn.

(i) Small $r$: in this limit both 
$\Pi _{2}d\Pi _{2}/d\Delta ^{2}$ in the numerator and 
$\Pi _{2}^{2}$ in the denominator of (\ref{fluctgapeq}) are  suppressed
by the small factor $(\nu /v_{F}\Lambda )^{2}$ relative to the other terms,
so we shall neglect the contribution from $\Pi _{2}$. 
In $d=2$ with the parabolic dispersion, $N_1=0$ identically, so this term vanishes altogether.
Then the leading
behavior of both $1+g\Pi _{0}$ and $g\Pi _{3}$ is $gN_{0}/2$; the leading
behavior of $d\Pi _{0}/d\Delta ^{2}$ is $N_{0}/2\Delta ^{2}$, and the
leading behavior of $d\Pi _{3}/d\Delta ^{2}$ is reduced compared to $%
N_{0}/2\Delta ^{2}$ by $r^{2}$. Thus, the numerator behaves as $%
(gN_{0})^{2}/4\Delta ^{2}$, and we factor the denominator as 
\begin{equation}
(1+g\Pi _{0}+g\Pi _{3})(1+g\Pi _{0}-g\Pi _{3})\simeq (gN_{0})^{2}1\times
r^{2}
\end{equation}%
identifying the two factors with amplitude and phase. That way, we find 
\begin{eqnarray}
{\frac{dS_{1}}{d\Delta ^{2}}}_{|small} &\approx &-{\frac{\Delta ^{d-1}}{%
v_{F}^{d}}}\int\limits_{0}^{1}{\frac{dv}{\pi }}\int\limits_{0}^{1}{\frac{%
S_{d-1}u^{d-1}du}{\pi ^{d}}}{\frac{1}{r^{2}}}  \label{GLfluctuations} \\
&\simeq &-{\frac{S_{d-1}\Delta ^{d-1}}{d\pi ^{d+1}v_{F}^{d}}}=-N_{0}{\frac{%
2^{d}}{\pi d}}\left( \frac{\Delta }{v_{F}\Lambda }\right) ^{d-1},  \nonumber
\end{eqnarray}%
with $u=\frac{v_{F}q}{2\Delta }$ and $v=\frac{\nu }{2\Delta }$, since, for a
spherical Fermi surface, 
\begin{equation}
N_{0}={\frac{S_{d-1}\Lambda ^{d-1}}{(2\pi )^{d}v_{F}}}.  \label{dos}
\end{equation}

(ii) Large $r$: The leading behavior of the individual terms in the
numerator and denominator of (\ref{fluctgapeq}) is%
\begin{eqnarray}
(1+g\Pi _{0})g{\frac{d\Pi _{0}}{d\Delta ^{2}}} &=&{\frac{(gN_{0})^{2}}{%
4\Delta ^{2}}}\left( \int\limits_{0}^{1}{\frac{d\mu (1-\mu ^{2})^{\frac{d-3}{%
2}}}{\kappa _{d}}}\ln r\right)  \\
&&\left( \int\limits_{0}^{1}{\frac{d\mu (1-\mu ^{2})^{\frac{d-3}{2}}}{\kappa
_{d}}}{\frac{1}{r^{2}}}\right)   \nonumber
\end{eqnarray}%
\begin{eqnarray}
g^{2}\Pi _{2}{\frac{d\Pi _{2}}{d\Delta ^{2}}} &{=}&{(gN_1)^2 \over 4\Delta ^2} \left( {\nu \over 2v_F\Lambda} \right) ^{2}\ln {\frac{v_{F}\Lambda }{%
\Delta }} \\
&&\left( \int\limits_{0}^{1}{\frac{d\mu (1-\mu ^{2})^{\frac{d-3}{2}}}{\kappa
_{d}}}{\frac{\ln r}{r^{2}}}\right)   \nonumber
\end{eqnarray}%
\begin{equation}
g^{2}\Pi _{3}{\frac{d\Pi _{3}}{d\Delta ^{2}}}={\frac{(gN_{0})^{2}}{4\Delta
^{2}}}\left( \int\limits_{0}^{1}{\frac{d\mu (1-\mu ^{2})^{\frac{d-3}{2}}}{%
\kappa _{d}}}{\frac{\ln r}{r^{2}}}\right) ^{2}
\end{equation}%
\begin{eqnarray}
(1+g\Pi _{0})^{2} &=&(gN_{0})^{2}\left( \int\limits_{0}^{1}{\frac{d\mu
(1-\mu ^{2})^{\frac{d-3}{2}}}{\kappa _{d}}}\ln r\right) ^{2} \\
(g\Pi _{2})^{2} &=&-(gN_{1})^{2}\left( {\frac{\nu }{2v_{F}\Lambda }}\right)
^{2}\ln ^{2}{\frac{v_{F}\Lambda }{\Delta }} \\
(g\Pi _{3})^{2} &=&(gN_{0})^{2}\left( \int\limits_{0}^{1}{\frac{d\mu (1-\mu
^{2})^{\frac{d-3}{2}}}{\kappa _{d}}}{\frac{\ln r}{2r^{2}}}\right) ^{2}.
\end{eqnarray}

We see that the leading contribution to both the numerator and denominator
comes from $\Pi _{0}$, so 
\begin{eqnarray}
{\frac{dS_{1}}{d\Delta ^{2}}} &\simeq &-\int^{\prime }{\frac{d\nu }{2\pi }}{%
\int\limits^{\Lambda \prime }}\left( {\frac{dq}{2\pi }}\right) ^{d}{\frac{g{%
\frac{d\Pi _{0}}{d\Delta ^{2}}}}{1+g\Pi _{0}}} \\
&\simeq &-{\frac{\Delta ^{d-1}S_{d-1}}{2v_{F}^{d}\pi ^{d+1}}}\int^{\prime
}u^{d-1}du\int^{\prime }dv\\
&&{\frac{\int\limits_{0}^{1}{\frac{d\mu (1-\mu ^{2})^{\frac{d-3}{2}}}{\kappa
_{d}}}{\frac{1}{r^{2}}}}{\int\limits_{0}^{1}{\frac{d\mu (1-\mu ^{2})^{\frac{%
d-3}{2}}}{\kappa _{d}}}\ln r}},  \nonumber
\end{eqnarray}%
where the prime means that we are integrating over large momenta and
frequencies compared to the gap. Now we have to consider separately the
regions $u<v$ and $u>v$. In the first region, we neglect the $u$ dependence
of $r$, so the $\mu $ (angular) integration becomes trivial, and we get 
\begin{eqnarray}
&&-{\frac{\Delta ^{d-1}S_{d-1}}{2v_{F}^{d}\pi ^{d+1}}}\int\limits_{1}^{%
\infty }{\frac{dv}{v^{2}\ln v}}\int\limits_{1}^{\min \left( v,{\frac{%
v_{F}\Lambda }{2\Delta }}\right) }u^{d-1}du  \label{u<v} \\
&\simeq &-{\frac{\Delta ^{d-1}S_{d-1}}{2v_{F}^{d}\pi ^{d+1}}}{\frac{1}{d-1}}{%
\frac{\left( {\frac{v_{F}\Lambda }{2\Delta }}\right) ^{d-1}}{\ln {\frac{%
v_{F}\Lambda }{\Delta }}}}=-{\frac{N_{0}}{\pi (d-1)\ln {\frac{v_{F}\Lambda }{%
\Delta }}}}.  \nonumber
\end{eqnarray}%
In the second region, we find, with logarithmic precision, 
\begin{equation}
\int\limits_{0}^{1}{\frac{d\mu (1-\mu ^{2})^{\frac{d-3}{2}}}{\kappa _{d}}}%
\ln 2\sqrt{v^{2}+(u\mu )^{2}}=\ln u,  \label{u>vlog}
\end{equation}%
and 
\begin{eqnarray}
&&\int\limits_{0}^{1}{\frac{d\mu (1-\mu ^{2})^{\frac{d-3}{2}}}{\kappa _{d}}}{%
\frac{1}{v^{2}+(u\mu )^{2}}}\approx {\frac{1}{\kappa _{d}uv}} \label{muint}
\end{eqnarray}%
so 
\begin{eqnarray}
&&-{\frac{\Delta ^{d-1}S_{d-1}}{2v_{F}^{d}\pi ^{d+1}}}\int\limits_{1}^{\frac{%
v_{F}\Lambda }{2\Delta }}u^{d-1}du\int\limits_{1}^{u}dv{\frac{1}{\kappa
_{d}uv\ln u}}  \label{high energy fluctuation correction} \\
&\simeq &-{\frac{\Delta ^{d-1}S_{d-1}}{2v_{F}^{d}\pi ^{d+1}}}{\frac{1}{%
(d-1)\kappa _{d}}}\left( {\frac{v_{F}\Lambda }{2\Delta }}\right) ^{d-1}=-{%
\frac{N_{0}}{\pi (d-1)\kappa _{d}}},  \nonumber
\end{eqnarray}%
so it is the dominant fluctuation contribution to the fluctuation correction
of the gap equation. We see that it is by factor $gN_{0}$ or $1/\ln
(v_{F}\Lambda /\Delta )$ smaller than the mean-field terms, and that it is
negative, so it decreases the value of the gap. This contribution will change
only the prefactor in the solution of the gap equation.  Other effects,
not considered in the present calculation, will also
make corrections of the same order\cite{Gorkov}.
Comparing formula (\ref{u<v}%
) to formula (\ref{high energy fluctuation correction}), we see that the
contribution from processes that scatter electrons along the Fermi surface
dominate other contributions by factor $\ln (v_{F}\Lambda /\Delta )$.

It is interesting to note that the result (\ref{high energy fluctuation
correction}) is small because of cancellation of large terms. Indeed,
returning to (\ref{fluctuation free energy}), we can write the fluctuation
correction to the gap equation (\ref{fluctgapeq}) as 
\begin{equation}
{\frac{dS_{1}}{d\Delta _{0}^{2}}}=-\beta V\int {\frac{d\nu }{2\pi }}%
\int\limits^{\Lambda }\left( {\frac{dq}{2\pi }}\right) ^{d}\mbox{Tr}\left[ g{%
\frac{d\mathbf{\Pi }}{d\Delta _{0}^{2}}}\left( \mathbf{1}+g\mathbf{\Pi }%
\right) ^{-1}\right] .
\end{equation}%
At high frequencies and momenta, the leading behavior of the two eigenvalues
of the matrix 
\begin{equation}
g{\frac{d\mathbf{\Pi }}{d\Delta _{0}^{2}}}\left( \mathbf{1}+g\mathbf{\Pi }%
\right) ^{-1},
\end{equation}%
that is, of the amplitude and phase mode, is $\pm f$ where 
\begin{eqnarray}
f &=&\int^{\prime }{\frac{d\nu }{2\pi }}{\int\limits^{\Lambda \prime }}%
\left( {\frac{dq}{2\pi }}\right) ^{d}{\frac{g\frac{d\Pi _{3}}{d\Delta ^{2}}}{%
1+g\Pi _{0}}}  \nonumber \\
&\simeq &{\frac{\Delta ^{d-1}}{v_{F}^{d}}}\int\limits_{1}^{\infty }{\frac{dv%
}{\pi }}\int\limits_{1}^{v_{F}\Lambda /2\Delta }{\frac{S_{d-1}u^{d-1}du}{\pi
^{d}}} \\
&&{\frac{\int\limits_{0}^{1}{\frac{d\mu (1-\mu ^{2})^{\frac{d-3}{2}}}{\kappa
_{d}}}{\frac{\ln r}{r^{2}}}}{\int\limits_{0}^{1}{\frac{d\mu (1-\mu ^{2})^{%
\frac{d-3}{2}}}{\kappa _{d}}}\ln r}}.  \nonumber
\end{eqnarray}%
Calculations similar to those leading to eqs (\ref{u<v})-(\ref{high energy
fluctuation correction}) show that, with logarithmic accuracy, 
\begin{eqnarray}
f &\simeq &{\frac{1}{v_{F}^{d}}}{\frac{\Delta ^{d-1}}{\pi ^{d+1}}}{\frac{%
S_{d-1}}{\kappa _{d}}}\int\limits_{1}^{v_{F}\Lambda /2\Delta
}u^{d-1}du\int\limits_{1}^{u}dv{\frac{{\frac{\ln v}{uv}}}{\ln u}} 
\nonumber \\
&\simeq &{\frac{1}{v_{F}^{d}}}{\frac{\Delta ^{d-1}}{\pi ^{d+1}}}{\frac{%
S_{d-1}}{\kappa _{d}}}{\frac{1}{2}}{\frac{1}{d-1}}\left( \frac{v_{F}\Lambda 
}{2\Delta }\right) ^{d-1}{\frac{\ln ^{2}{\frac{v_{F}\Lambda }{\Delta }}}{\ln 
{\frac{v_{F}\Lambda }{\Delta }}}}  \nonumber \\
&\simeq &{\frac{N_{0}}{\pi (d-1)\kappa _{d}}}\ln {\frac{v_{F}\Lambda }{%
\Delta }},
\end{eqnarray}

Thus each of the amplitude and phase mode gives a correction apparently
large enough to call the BCS approximation into question, but the two terms
cancel in the trace.

\section{Summary and Conclusions}

We have used a functional integral formulation to study fluctuation
corrections to the weak coupling BCS mean field expression for the
superconducting gap amplitude. We find that in both two and three dimensions
and in the weak coupling limit, the correction is smaller by the
weak-coupling factor $gN_{0}$ or $1/\ln (v_{F}\Lambda /\Delta )$ than the
mean-field terms. This correction comes from energies and momenta large
compared to the mean-field gap, and, more specifically, from the region in
the phase space where the electrons are scattered nearly parallel to the
Fermi surface. We note that individually, both the contribution from the
amplitude mode and the contribution from the phase mode diverge
logarithmically in this region, but the divergencies have opposite sign, and
thus cancel when we take the trace.

Acknowledgements: A. J. M was supported by NSF-DMR-0431350. \v{S}. K.
acknowledges the LANL DR Project 200153 and the Department of Energy, under
contract W-7405-ENG-36, and A.I.L. NSF 0439026. We thank L. B. Ioffe for
helpful discussions.

\end{document}